\documentclass[prl,aps,twocolumn,10pt,showpacs,groupedaddress,superscriptaddress,floatfix,notitlepage]{revtex4-2}
\usepackage{amsmath,amsfonts,amssymb,graphics,graphicx,epsfig,color,times, braket}
\usepackage{color}
\usepackage{bbm, dsfont}
\usepackage{subfigure}
\usepackage{hyperref} 
\usepackage{mathrsfs}
\usepackage{verbatim} 
\begin{document}

\title{Informational Mpemba Effect for Fast State Purification in Non-Hermitian System}

\author{C. G. Feyisa}
\email{chimdessagashu@gmail.com}
\affiliation{Institute of Atomic and Molecular Sciences, Academia Sinica, Taipei 10617, Taiwan}
\affiliation{Molecular Science and Technology Program, Taiwan International Graduate Program, Academia Sinica, Taipei 10617, Taiwan}
\affiliation{Department of Physics, National Central University, Taoyuan 320317, Taiwan}

\author{Huan-Yu Ku}
\affiliation{Department of Physics, National Taiwan Normal University, Taipei 11677, Taiwan}

\author{J.-S. You}
\affiliation{Department of Physics, National Taiwan Normal University, Taipei 11677, Taiwan}

\author{H. H. Jen}
\email{sappyjen@gmail.com}
\affiliation{Institute of Atomic and Molecular Sciences, Academia Sinica, Taipei 10617, Taiwan}
\affiliation{Molecular Science and Technology Program, Taiwan International Graduate Program, Academia Sinica, Taipei 10617, Taiwan}
\affiliation{Department of Physics, National Central University, Taoyuan 320317, Taiwan}
\affiliation{Physics Division, National Center for Theoretical Sciences, Taipei 10617, Taiwan}

\date{\today}
\renewcommand{\r}{\mathbf{r}}
\newcommand{\f}{\mathbf{f}}
\renewcommand{\k}{\mathbf{k}}
\def\p{\mathbf{p}}
\def\q{\mathbf{q}}
\def\bea{\begin{eqnarray}}
\def\eea{\end{eqnarray}}
\def\ba{\begin{array}}
\def\ea{\end{array}}
\def\bdm{\begin{displaymath}}
\def\edm{\end{displaymath}}
\def\red{\color{red}}
\pacs{}
\begin{abstract}
Quantum systems are inherently fragile to environmental fluctuations or decoherence, limiting their advantages in applications of quantum information and quantum computation. State purification offers a route to recover the purity of system under noisy conditions. Here, we demonstrate a rapid purification of initially mixed states by harnessing collective reservoir engineering in driven non-Hermitian qubit systems, together with multipartite entanglement generation in larger systems. We show that the onset of efficient purification-assisted entanglement generation is dictated by the degeneracy of collective subradiant modes, rather than by exceptional points. Moreover, the system dynamics manifests an informational Mpemba effect, i.e., a more mixed initial state reaches its steady state with unit purity at a faster rate, resembling the conventional Mpemba effect where a hotter system cools more rapidly. These results reveal a unique advantage of driven non-Hermitian quantum systems with engineered collective dissipation, enabling enhanced purification efficiency and offering new opportunities for relaxation control of non-Hermitian systems and quantum state preparations.
\end{abstract}
\maketitle
{\it Introduction}--Protecting quantum states against decoherence is a central challenge in quantum information processing and quantum computation \cite{Nielsen2000}. To mitigate the inevitable effects of fluctuations in open quantum systems, quantum state purification \cite{Bennett1996_1, Bennett1996_2, Bennet1996_3} provides a principled route to restore a system’s purity under noisy environments. Such purification may be achieved through continuous measurements \cite{Wiseman2010} or quantum feedback control \cite{Griffith2007, Ralph2011, Shabani2013}, which in turn enables entanglement purification \cite{Horodecki2009}, magic state distillation \cite{Knill2004, Bravyi2005, Ye2023, Gupta2024, Rodriguez2025}, and more generally quantum resource distillation \cite{Chitambar2019, Nery2020, Ku2022, Ku2023, Ku2023_2, Hsieh2025, Stratton2025}. These developments collectively pave the way toward universal and scalable fault-tolerant quantum computation \cite{Shor1996, Knill1996, Gottesman1998}.

Beyond state protection, continuously monitoring quantum systems allows access to measurement-induced phase transitions \cite{Skinner2019, Gullans2020, Gopalakrishnan2021, Ippoliti2021, Mochizuki2025, Poboiko2025, Yokomizo2025, Delmonte2025, Paviglianiti2025, Li2025}, where entanglement displays either volume-law or area-law scaling and purification transition emerges depending on the rates of quantum measurement. Meanwhile, an even richer interplay has recently been explored between engineered dissipation \cite{Verstraete2009, Harrington2022} and higher-order exceptional points (EPs) in non-Hermitian qubits \cite{El-Ganainy2018, Ozdemir2019, Ashida2020}, leading to accelerated \cite{Li2023, Feyisa2025_1, Feyisa2025_2} and amplified entanglement generation \cite{Hotter2025}. These effects are particularly relevant for fast quantum operations and high-fidelity state preparation \cite{Karmakar2025}. However, achieving robust and reliable generation of purified entangled states remains challenging, highlighting the need for controllable approaches, most notably through collective reservoir engineering \cite{Harrington2022, Hotter2025} to support their efficient preparation.

In this Letter, we harness collective reservoir engineering in non-Hermitian qubit systems to accelerate the purification of completely mixed states initially in a driven-dissipative platform. Notably, collective dissipation creates a hierarchy of decay rates that separates the eigenmodes into fast-decaying superradiant sectors and long-lived subradiant sectors. The former host exceptional points (EPs), whereas the latter exhibit subradiant mode degeneracy at a critical driving strength, and tuning the system close to or away from this critical point slows or accelerates the relaxation toward an maximmally entangled state in two qubits and multipartite entangled states in larger systems. More intriguingly, the system dynamics exhibit an informational Mpemba effect (ME) \cite{Hsieh2025}, whereby a more mixed initial state reaches its steady state with unit purity at a faster rate and with a higher success probability. This mirrors the conventional ME \cite{Mpemba1969, Carollo2021, Chatterjee2023, Chatterjee2024, Chalas2024, Rylands2024, Joshi2024, Shapira2024, Turkeshi2025, Ares2025, Zhang2025, Nava2025, Ma2025,  Teza2026}, in which a hotter initial state cools more rapidly. This anomalous relaxation behavior underscores a distinct advantage of driven-dissipative systems with engineered collective dissipation, enabling enhanced efficiency in quantum engineering and state preparation.
 
\begin{figure}[b]
	\centering	\includegraphics[width=0.48\textwidth]{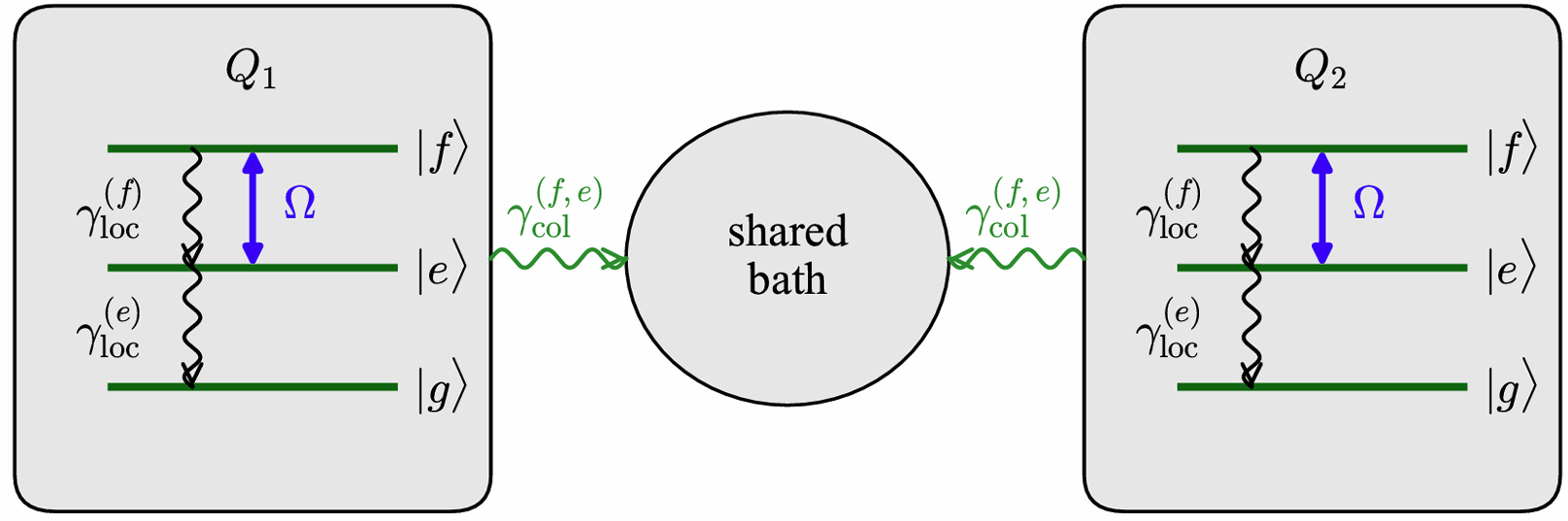}
	\caption{A schematic plot of two non-Hermitian qubits under reservoir engineering. Exemplary transmon qubits $Q_{1,2}$ are driven by a coherent drive $\Omega$ in the effective subspace of $|e\rangle$ and $|f\rangle$ with local decay rates $\gamma_{\rm loc}^{(e)}\gg\gamma_{\rm loc}^{(f)}$, effectively constructing a non-Hermitian setup strongly monitored by the state $|g\rangle$. A shared bath presents a pathway to engineer the collective dissipations $\gamma_{\rm col}^{(f,e)}$ among $Q_{1,2}$.}\label{fig1}
\end{figure}

{\it Model}--We consider a generic non-Hermitian qubit system exemplified in transmon superconducting circuits \cite{Naghiloo2019, Kjaergaard2020, Chen2021} in Fig. \ref{fig1}. These qubits can be constructed within the manifold of a three-level system, the ground state $|g\rangle$, the first excited state $|e\rangle$, and the second excited state $|f\rangle$. A two-level subspace of a qubit $|e\rangle$ and $|f\rangle$ effectively forms when the ground state is treated as a reservoir with a dominant decay channel, monitoring the effective two-level qubit system. This can be achieved by utilizing impedance-mismatching element to amplify or suppress electromagnetic radiation mode in a three-dimensional microwave cavity \cite{Naghiloo2019, Chen2021}.

The system's Hamiltonian can be written as
\bea
H_{\rm sys} =&& \sum_{j=1}^N \left[\Delta_j \hat L_j^{f,\dag} \hat L_j^f +\Omega(\hat L_j^{f, \dag} + \hat L_j^f)\right]+J\sum_{j\neq k}^N \hat L_j^{f,\dag} \hat L_k^f,\nonumber\\
\eea
where $\Omega$ denotes the Rabi frequency of the coherent driving field with a detuning $\Delta_j$, and $J$ indicates an inter-qubit coupling constant. The master equation for a density matrix $\rho$ becomes ($\hbar=1$) \cite{Li2023}
\begin{eqnarray}
\frac{d}{dt}\rho(t)=-i\left(H_{\rm eff}\rho-\rho H_{\rm eff}^{\dagger}\right)
+\sum_{j,k=1}^N\Gamma_{jk}^{f}\hat L_f^{j}\rho\hat L_f^{k\dagger},\label{rho}
\end{eqnarray}
in which $ H_{\rm eff}=H_{\rm sys}-(i/2)\sum_{\alpha\in\{e,f\}}\sum_{j,k=1}^N\Gamma_{jk}^{\alpha}\hat L_\alpha^{j\dagger}\hat L_\alpha^{k}.$ The collective dissipation rates $\Gamma^\alpha_{j,k}\equiv \gamma^\alpha_{\rm loc}\delta_{jk}+\gamma_{j,k}^{\rm {\alpha}}$ involve a local intrinsic decay rate $\gamma^\alpha_{\rm loc}+\gamma^{\alpha}_{j,j}=\gamma_\alpha$, determined by a total decay rates of $\ket{e}$, $\ket{f}$, and the associated collective ones $\gamma^{e}_{j,k}$ and $\gamma^{f}_{j,k}$ among qubits with $\hat L_{j}^e\equiv | g \rangle_j \langle e|$ and $\hat L_{j}^f\equiv | e \rangle_j \langle f|$. A ratio of $\eta_\alpha\equiv\gamma_{j,k}^{\alpha}/\gamma_\alpha$ quantifies the relative strength between these two decaying channels, highlighting the role of collective dissipation.

Throughout the paper, we consider $\Delta_j=0$ and $J=0$, assume $\gamma_e \gg \gamma_f$, and focus on purely driven-dissipative non-Hermitian systems with uniform all-to-all collective decay rates, $\gamma_{j,k}^{e}=\gamma_c$, where $\eta\equiv\gamma_c/\gamma_e$. Our results should also apply to the case of a finite $J$, since it only lifts the EPs and pushes the degeneracy points toward a larger $\Omega$ \cite{SM}, irrelevant to the anomalous relaxations attributed to ME or informational ME we discuss here. By removing the jump terms in Eq. (\ref{rho}), we focus on the effective subspace of the system in the non-Hermitian regime and reconstruct $\rho(t)\rightarrow\rho(t)/\textrm{Tr}[\rho(t)]$, valid for open quantum systems upon post-selection \cite{Ashida2020}. In the End Matter, we discuss the effect of quantum jumps and provide trajectory-level analysis, reaffirming the results in this work on fast state purification.

{\it Informational Mpemba effect for single qubit}--First we investigate the properties of a non-Hermitian single qubit, where we find both the intriguing informational ME and conventional ME in distinct parameter regimes. For mixed systems, relaxation can occur toward either high-entropy, maximally mixed stationary states or low-entropy pure steady states. The former corresponds to conventional dissipative mixing, which commonly arises in thermalization and decoherence processes, whereas the latter is associated with purification dynamics, which plays an essential role in resourceful quantum state preparation.

\begin{figure}[t]
	\centering	\includegraphics[width=0.48\textwidth]{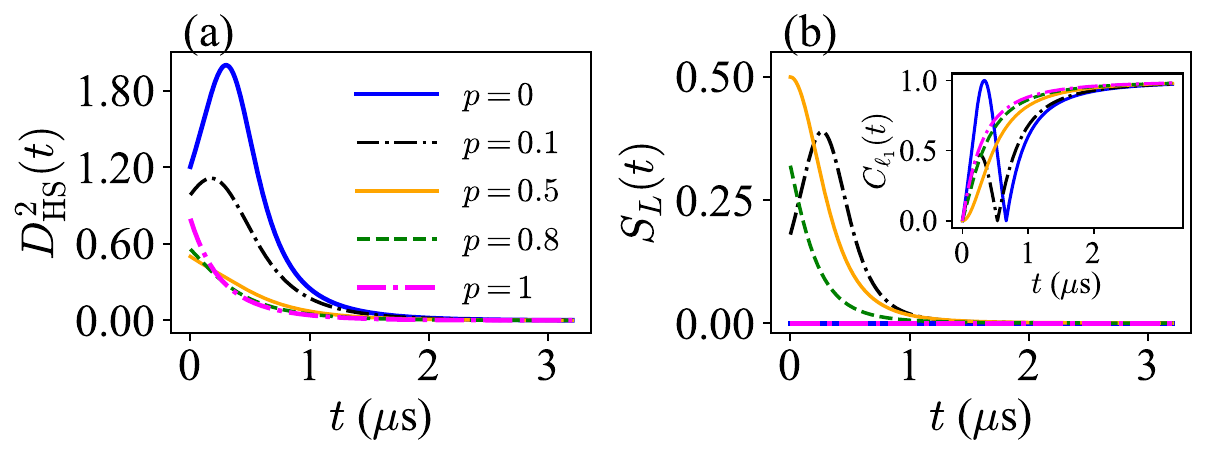}
	\caption{Relaxation dynamics of a single non-Hermitian qubit at the EP ($\Omega=\gamma_e/4$, $\gamma_e=6~{\rm rad}/\mu{\rm s}$), starting from the diagonal state $\rho(0)=p\ket{f}\!\bra{f}+(1-p)\ket{e}\!\bra{e}$ with $0\leq p\leq1$. 
		(a) Time evolution of the $D_{\rm HS}^2(t)$ between $\rho(t)$ and the steady state $\ket{\psi_{\rm ss}}=(\ket{f}-i\ket{e})/\sqrt{2}$, which is the unique and dominant eigenstate of the effective non-Hermitian dynamics at the EP. (b) Linear entropy $S_L(t)$ illustrates the purification dynamics toward the steady state, featuring an informational ME. The inset in Fig.~(b) shows the transient buildup of quantum coherence, quantified by the $\ell_1$-norm $C_{\ell_1}(t)$, during the anomalous relaxation dynamics. 
	}
	\label{fig2}
\end{figure}

In Fig. \ref{fig2}(a), we calculate the Hilbert-Schmidt (HS) distance $D_{\rm HS}(t)\equiv\sqrt{{\rm Tr}[\rho(t)-\rho_{ss}]^2}$ \cite{Ares2025, Teza2026} and plot $D^2_{\rm HS}$ for better comparisons, which quantifies how far the system evolves toward the steady-state density matrix $\rho_{ss}={\rm lim}_{t\rightarrow\infty} \rho(t)$. $\rho_{ss}$ is always pure when $\Omega\leq \gamma_e/4$ in the $\mathcal{PT}$ (parity and time symmetry)-broken regime. Intriguingly, only under a certain parameter spaces $p>0.5$ from an initial diagonal state, the system farther from equilibrium evolves into the steady state faster, showing a clear feature of ME~\cite{Lu2017, Klich2019, Carollo2021}.

\begin{figure*}[t]
	\centering
    \includegraphics[width=\textwidth]{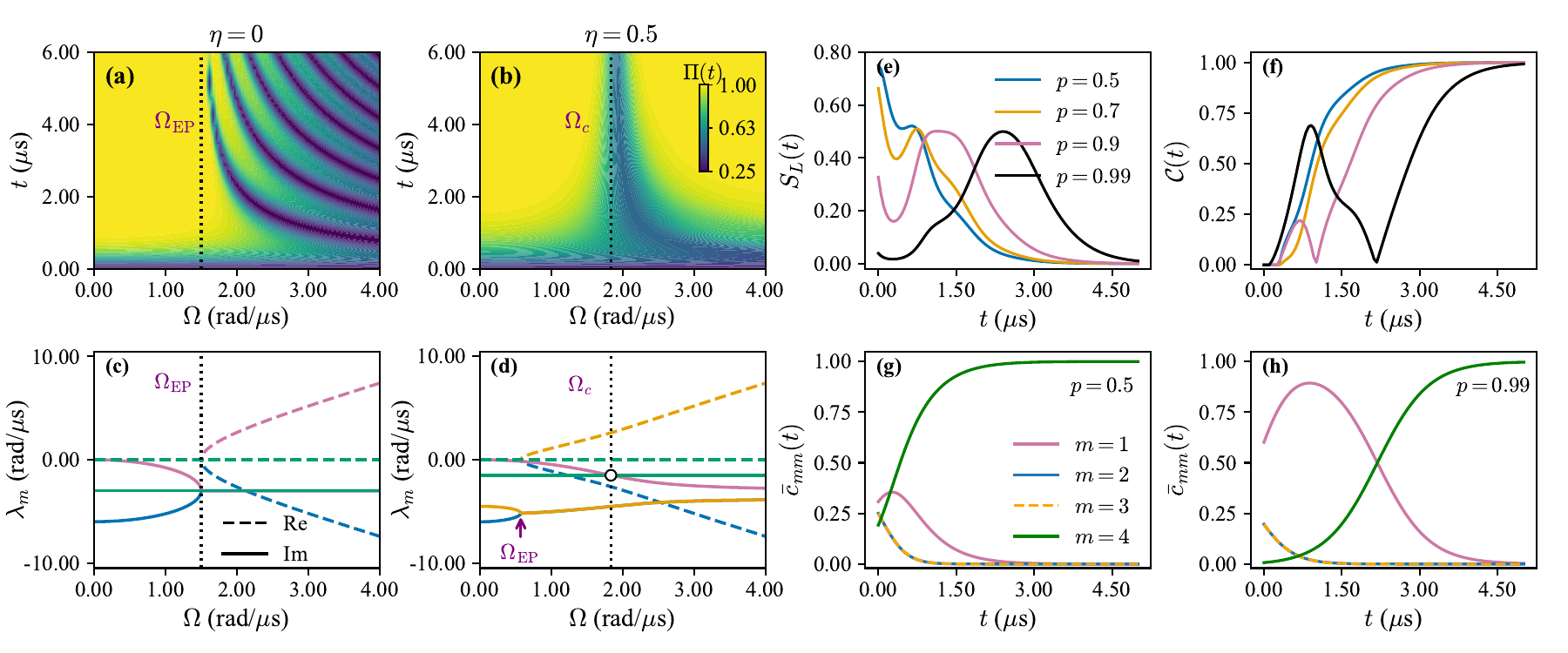}
	\caption{Purification dynamics and spectral properties of two non-Hermitian qubits initialized in $\rho(0)=\big[(1-p)\ket{e}\!\bra{e}+p\ket{f}\!\bra{f}\big]^{\otimes 2}$. (a,b) Purity dynamics $\Pi(t)$ from the maximally mixed state ($p=0.5$) as a function of time and driving strength for (a) $\eta=0$ and (b) $\eta=0.5$. (c,d) Corresponding complex eigenvalue spectrum $\lambda_m$ versus $\Omega$, where solid (dashed) curves denote $\mathrm{Im}[\lambda_m]$ ($\mathrm{Re}[\lambda_m]$). The modes $\lambda_{1,2,3}$ (magenta, blue, orange) belong to the symmetric sector spanned by $\ket{ff}$, $\ket{ee}$, and $\ket{S}\equiv(\ket{ef}+\ket{fe})/\sqrt{2}$, whereas $\lambda_4$ (green) corresponds to the antisymmetric state $\ket{A}\equiv(\ket{ef}-\ket{fe})/\sqrt{2}$. The vertical dotted lines indicate the EP at $\Omega_{\rm EP}=\gamma_e/4$ for $\eta=0$ and the subradiant-mode degeneracy at $\Omega_c=\gamma_e/(2\sqrt{2})\sqrt{1-\eta^2}$ for $\eta=0.5$, where the slowest-decaying mode switches between the symmetric and antisymmetric sectors. (e,f) Linear entropy $S_L(t)$ and concurrence $\mathcal{C}(t)$ for $\eta=0.5$, $\Omega=3~{\rm rad}/\mu{\rm s}$, and initial states $p=0.5$ (blue), $0.7$ (orange), $0.9$ (magenta), and $0.99$ (black). The dynamics exhibits an information Mpemba effect, where initially more mixed states purify and reach maximal entanglement faster under non-Hermitian dynamics. (g,h) Eigenmode decomposition for $p=0.5$ and $p=0.99$, respectively, with colors matching the eigenvalues $\lambda_{1,2,3,4}$ shown in (c,d).}
	\label{fig3}
\end{figure*}
 
We note that $D_{\rm HS}^2=\Pi(t)+\textrm{Tr}[\rho_{ss}^2]-2\textrm{Tr}[\rho(t)\rho_{ss}]$, where the purity of the system is $\Pi(t)\equiv \textrm{Tr}[\rho^2(t)]$, and the last term, if $\rho_{ss}$ is pure, indicates the fidelity of the system to the steady state up to a factor of $2$. If the stationary state is maximally mixed, $\rho_{ss}=\mathds{I}_d/d$, where $\mathds{I}_d$ denotes the identity operator acting on a Hilbert space of dimension $d$, the $D^2_{\rm HS}$ follows directly as $\Pi(t)-d^{-1}$, showing a direct analogy between HS distance and the purity up to a constant determined by the system's dimensionality \cite{Hsieh2025}. In Fig. \ref{fig2}(b), we plot the linear entropy $S_L(t)\equiv 1 -\Pi(t)$ instead for clarity, which measures the mixedness of the system and also reveals additional features that are not visible in the HS distance. In particular, two distinct dynamical regimes emerge depending on the initial population. For $0.5 < p < 1$, the entropy decreases monotonically, indicating direct purification toward the stationary pure state, where less mixed states relax faster. Interestingly, when $p*<p\leq 0.5$ with $p*$ close to $0$, $S_L$ shows non-monotonic behavior and facilitated state purification when the system is more mixed, featuring an informational ME. This effect emerges because more mixed initial states possess a larger overlap with the slowest-decaying eigenmode than purer initial states and therefore relax more rapidly toward the steady state.

The relaxation process involves two steps: The state first evolves toward a more mixed configuration before relaxing into the pure steady state, reflecting a Pontus Mpemba effect \cite{Nava2025}. The initial entropy growth is analogous to effective heating, followed by purification through entropy reduction, corresponding to cooling to stationary state~\cite{Pichler2015}. While anomalous relaxation in purity persists even for $\Omega=0$~\cite{SM}, coherent driving enables resource-efficient quantum-state preparation through the informational Mpemba effect, as discussed below in connection with exceptional points and subradiant-mode degeneracies.

Next, we investigate the system's coherence by using the most general quantifier, the $\ell_1$ norm of coherence $C_{\ell_1}(t)=2|\langle e|\rho(t)|f\rangle|$ in the single qubit case \cite{Baumgratz2014}. Notably, $C_{\ell_1}(t)$ in Fig. \ref{fig2}(b) resembles a two-step process as in $S_L(t)$ in a range of $0<p\leq 0.5$ and showcases a speedup in reaching the maximal coherence $C_{\ell_1}(t)=1$ near an exceptional point when the initial state is more mixed. On the other hand, an accelerated development of quantum coherence in $C_{\ell_1}(t)$ at $p>0.5$ coincides with the ME in $D_{\rm HS}$, illustrating another class of ME-assisted buildup of quantum resources \cite{Chitambar2019, Summer2025} and even surpassing the speed of the initial coherent state with maximal $C_{\ell_1}(0)=1$ in the non-Hermitian regime~\cite{SM}.

{\it Fast state purification with reservoir engineering}--To further explore state purification in a system of multiple qubits beyond noninteracting ones, in Fig. \ref{fig3} we harness the reservoir engineering of collective decays and utilize informational ME for fast state purification process. Starting from the maximally mixed states, we show the time-evolving purity $\Pi(t)$ with a dependence of collective dissipation ratio $\eta$ and $\Omega$. 

To obtain $\Pi(t)$, we construct $\rho(t)$ from the left and right eigenstates of $H_{\rm eff}=H_{\rm sys}-i/2\sum_{j,k=1}^2\Gamma_{j,k}\hat L_j^{e,\dag}\hat L_k^e$, 
\bea
H_{\rm eff}\ket{\psi_n^R}&&=\lambda_n\ket{\psi_n^R}, \\
H_{\rm eff}^\dag\ket{\psi_n^L}&&= \lambda_n^*\ket{\psi_n^L},
\eea
where $\lambda_n$ denotes the eigenvalues and biorthogonality relation preserves as $\bra{\psi_m^L}\psi_n^R\rangle=\delta_{mn}$. The density matrix can then be constructed as 
\bea
\rho(t)&&\propto\sum_{m,n=1}^4c_{mn}(0)e^{-i(\lambda_m-\lambda_n^*)t}\ket{\psi_m^R}\bra{\psi_n^R},\label{rho2} \\
c_{mn}(0)&&\equiv \bra{\psi_n^L} \rho(t=0)\ket{\psi_m^L},
\eea
where $\rho(t)$ is up to renormalization to preserve $\textrm{Tr}[\rho(t)]=1$ as in continuously monitored systems \cite{Daley2014}. From the above, the purity can be calculated as 
\bea
\Pi(t)=\frac{\displaystyle
\sum_{m,n,k,l=1}^4c_{mn}(t)c_{kl}(t)
\langle\psi_n^R|\psi_k^R\rangle
\langle\psi_l^R|\psi_m^R\rangle}
{\displaystyle\left[\sum_{m,n=1}^4
c_{mn}(t)\langle\psi_n^R|\psi_m^R\rangle
\right]^2},
\eea
where $c_{mn}(t)=c_{mn}(0)e^{-i(\lambda_m-\lambda_n^*)t}$.
%
%

Figure \ref{fig3}(a) shows a reference of $\Pi(t)$ for two noninteracting qubits, where coherent states can be purified in the $\mathcal{PT}$-broken regime, as exemplified in Fig. \ref{fig2} for a single qubit. In $\mathcal{PT}$-symmetric regime at $\Omega>\gamma_e/4$, on the other hand, fast oscillations emerge in $\Pi(t)$ due to the beatings in frequencies determined by $\textrm{Re}(\lambda_1)-\textrm{Re}(\lambda_2)$, as shown in Figs.~\ref{fig3}(a) and \ref{fig3}(c). In Figs. \ref{fig3}(b), two regimes of $\Pi(t)$ can be identified and separated by the degeneracies of complex eigenvalues. This similar crossing in eigenmodes as a phase transition is also observed in the relaxation dynamics of proposed four-state colloidal system \cite{Teza2023}. Across the degeneracy points in Figs. \ref{fig3}(d), the dominant subradiant eigenmodes, min$\{|\textrm{Im}(\lambda_m)|\}$, transition from $\ket{ff}$ at $\Omega=0$ to $\ket{A}=(\ket{ef}-\ket{fe})/\sqrt{2}$, manifesting a generation of maximal bipartite entanglement while being purified. Notably, as $\eta$ increases, the regime for purified state $\ket{A}$ arises in a faster rate, showing an enhanced state purification along with a creation of maximal entanglement. This attributes to the enlarged dissipative gap of $|\mathrm{Im}(\lambda_1)-\mathrm{Im}(\lambda_4)|$ for an increasing $\eta$, which removes the mode degeneracies at a larger $\Omega$ and therefore, facilitates the purification process under driven-dissipative conditions. We will discuss and explore more on the dissipative gap below. 

Furthermore, as demonstrated in Figs. \ref{fig3} (e-h) when $p$ varies in initially diagonal product states, we uncover informational ME-accelerated quantum-state purification. That is, the more mixed states with larger $S_L$, the faster they relax toward purified and entangled states. This accelerated state purification can be seen in Fig. \ref{fig3}(e), where a more purified initial state relaxes with a delayed heating stage as in the two-step process in Fig. \ref{fig2}(b). It is the anomaly of mitigated purification at an intermediate time that leads to the informational ME. Similar relaxation behavior also appears in the concurrence $\mathcal{C}(t)=\textrm{max}\{0,a_1-a_2-a_3-a_4\}$ in Fig. \ref{fig3}(f), where $a_m$ indicates the eigenvalues of the Hermitian matrix $\sqrt{\sqrt{\rho(t)}\tilde\rho(t)\sqrt{\rho(t)}}$ with $\tilde\rho(t)\equiv(\sigma_y\otimes\sigma_y)\rho^*(t)(\sigma_y\otimes\sigma_y)$ and the Pauli-$y$ matrix $\sigma_y$ \cite{Wootters1998, Verstraete2001, Feyisa2025_2}. This showcases a faster generation of maximal bipartite entanglement when the system is more mixed. 

By decomposing the system with time-dependent mode overlaps $c_{mn}(t)$ in Eq. (\ref{rho2}), in Figs. \ref{fig3}(g, h) we show the respective weights of each mode as time evolves, that is $\bar c_{mm}(t)\equiv c_{mm}(t)/\sum_m c_{mm}(t)$. As the system is more mixed initially at $p=0.5$, the dominant subradiant mode with a weight of $c_{44}(t)$ rapidly grows to its maximum, while suppressing the next dominating one of $c_{11}(t)$. As a comparison when $p$ increases initially, approaching the pure state limit, the weight of $c_{44}(t)$ arises slowly and reaches its maximum only after $c_{11}(t)$ is significantly reduced. This can be explained if we focus on the two dominant subradiant modes $\lambda_1$ and $\lambda_4$ as shown in Fig. \ref{fig3}(d) when crossing beyond the degeneracy points. The density matrix can be approximated as \cite{SM}
\bea
\rho(t)\approx \frac{c_{11}(0)e^{2{\rm Im}\lambda_1t}\ket{\psi_1^R}\bra{\psi_1^R}+c_{44}(0)e^{2{\rm Im}\lambda_4t}\ket{\psi_4^R}\bra{\psi_4^R}}{c_{11}(0)\langle\psi_1^R|\psi_1^R\rangle e^{2{\rm Im}\lambda_1t}+c_{44}(0)e^{2{\rm Im}\lambda_4t}},\nonumber\\
\eea
where $c_{14}=0$ due to the orthogonality between symmetric and antisymmetric sectors. This leads to $\Pi(t)=[\varepsilon(t)^2+1]/[\varepsilon(t)+1]^2$, where $\varepsilon(t)\equiv [c_{11}(0)\langle\psi_1^R|\psi_1^R\rangle/c_{44}(0)]e^{-2({\rm Im}\lambda_4-{\rm Im}\lambda_1)t}$ and $c_{44}(0)=p(1-p)$ \cite{SM}. When $p\rightarrow 0.5$, the more mixed the system, $c_{44}(0)$ becomes the largest, which reduces $\varepsilon(t)$ at $t=0$, rendering the system an advantage to be purified faster. This shows the essential role of mode overlap at initial time, which determines the fate and speed of the system to relax toward the steady state. Notably, the speed of relaxation also depends on the dissipative gap $|{\rm Im}\lambda_4-{\rm Im}\lambda_1|$ as shown in $\varepsilon(t)$. This explains the gap opening in Fig. \ref{fig3}(d) that leads to a speedup in state purification. We note that ordinary eigenmode filtering is governed by the distinct imaginary parts of the non-Hermitian spectrum, which determine the asymptotic state and the relaxation timescale, whereas the informational ME arises from different initial-state overlaps with the same eigenvalue spectrum, leading to a reversal of the relaxation ordering in the purification dynamics.

\begin{figure}[t] 
	\includegraphics[width=0.48\textwidth]{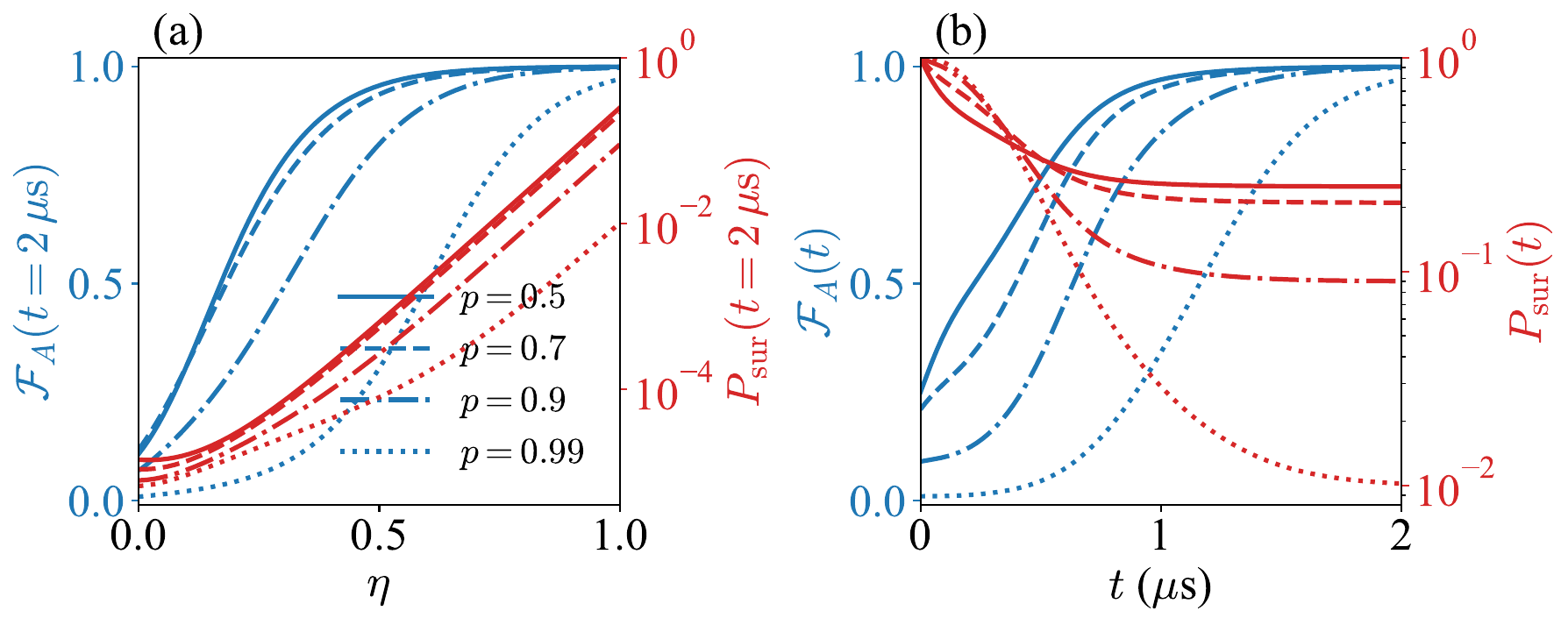}
	\caption{Fidelity and survival probability of two non-Hermitian qubits. (a) Bell-state fidelity, $\mathcal{F}_A(t)=\langle A|\rho(t)|A\rangle$ (blue), with $\ket{A}=(\ket{ef}-\ket{fe})/\sqrt{2}$, and survival probability, $P_{\rm sur}(t)={\rm Tr}[\rho(t)]$ (red), for varying collective dissipation $\eta$ at $t=2~\mu{\rm s}$. (b) Corresponding time evolution for the initial state $\rho(0)=\big[(1-p)\ket{e}\bra{e}+p\ket{f}\bra{f}\big]^{\otimes2}$ with $p=0.5$ (solid), $0.7$ (dashed), $0.9$ (dash-dotted), and $0.99$ (dotted), for $\eta=1$ and $\Omega=3~{\rm rad}/\mu{\rm s}$. Increasing the degree of collective dissipation enhances the survival probability by protecting the long-lived subradiant eigenmode while preserving a high fidelity with the entangled Bell state.}
	\label{fig4}
\end{figure}

In Fig.~\ref{fig4}, we further analyze the no-jump survival probability and the fidelity to the target state. As $\eta$ increases in Fig. \ref{fig4}(a), both of the properties are enhanced, indicating a rapid purification rate and significant survival probability due to the enlarged dissipative gap and the suppression of the decay rate of the antisymmetric eigenmode. This purification efficiency is further boosted for more mixed initial states due to their larger overlap with the protected subradiant mode, enabling even faster preparation of high-fidelity entangled states and demonstrating the operational usefulness of the information ME as shown in Fig.~\ref{fig4}(b).

\textit{Multiqubit case}---Our results also apply to multiqubit systems under driven-dissipative conditions. In the weak-driving regime, the informational Mpemba effect occurs at $p<0.5$ and accelerates the relaxation of mixed initial states toward a unique multipartite-entangled steady state arising from drive-induced hybridization within the symmetric Dicke sector. Under strong driving, however, a subradiant-mode degeneracy at the parity-dependent critical strengths $\Omega_c^{(\mathrm{even})}$ and $\Omega_c^{(\mathrm{odd})}$ switches the slowest-decaying modes to the lowest-spin Dicke sectors for even and odd $N$, respectively \cite{SM}. We analyze these dynamics in the Dicke basis $|j,m\rangle$ \cite{Dicke1954}, where $j=0,1/2,\ldots,N/2$ and $m=-j,\ldots,j$, with degeneracy ${\rm g}_j=N!(2j+1)/[(N/2+j+1)!(N/2-j)!]$. For even $N$ and $\Omega>\Omega_c^{(\mathrm{even})}$, the slowest-decaying sector comprises ${\rm g}_0$ degenerate $j=0$ singlets with an intrinsic $\Pi_{\rm even}\rightarrow1/{\rm g}_0$. The asymptotic state is therefore a pure entangled state for $N=2$ (${\rm g}_0=1$) and a statistical mixture of singlets for larger even $N$ (${\rm g}_0>1$), while the informational Mpemba effect persists \cite{SM}.

In contrast, for odd $N$ and $\Omega>\Omega_c^{(\mathrm{odd})}$, the dynamics is confined to the ${\rm g}_{1/2}$-fold degenerate $j=1/2$ manifold, with $\Pi_{\rm odd}(t)\approx\Pi_{1/2}(t)/{\rm g}_{1/2}$. Each doublet behaves as an effective $\mathcal{PT}$-symmetric qubit and undergoes coherent dynamics with $\Pi_{1/2}(t)=1-{2[1+\gamma_e^2(1-\cos\chi t)/(4\chi^2)]^2}^{-1}$ and period $2\pi/\chi$, where $\chi=\sqrt{4\Omega^2-\gamma_e^2/4}$. These coherent oscillations persist in the many-body dynamics, while the degeneracy ${\rm g}_{1/2}$ sets the overall purity scale. Despite these parity-dependent dynamics, both even- and odd-$N$ systems exhibit the universal large-$N$ scaling $\Pi\propto N^{3/2}\Pi(0)$ \cite{SM}, where $\Pi(0)=2^{-N}$. This $N^{3/2}$ enhancement originates from the asymptotic confinement of the many-body dynamics to the lowest-spin sectors, which host decoherence-free and coherent subspaces accessible through relaxation control of non-Hermitian dynamics via subradiant-mode degeneracy and the informational Mpemba effect.

{\it Experimental feasibility}--Our results are readily implementable across several platforms, including circuit QED systems \cite{Blais2021}, superconducting qubits with all-to-all couplings \cite{Roy2020, Zhou2023, Pita-Vidal2025} or bath engineering \cite{Harrington2019}, semiconductor quantum dots \cite{Chatterjee2023}, and trapped ions with tunable decay channels \cite{Zhang2025}. These platforms are also scalable for exploring anomalous relaxation dynamics and state purification enabled by engineered dissipation \cite{Verstraete2009, Harrington2022}. We therefore anticipate broad opportunities for observing novel nonequilibrium phenomena, as well as developing new approaches for quantum state preparation and resource-efficient quantum operations in state-of-the-art quantum platforms.

In the End Matter, we extend our analysis to the full Lindblad evolution. The informational ME persists in the presence of quantum jumps, with initially less pure states relaxing faster toward the high-purity steady state. Increasing the jump rate $\gamma_f$ accelerates relaxation, whereas smaller $\gamma_f$ yields higher steady-state purity. Our results therefore demonstrate rapid and robust quantum-state purification under realistic experimental conditions.

{\it Conclusion and outlook}--In this work, we uncover an informational ME-assisted state purification in open quantum systems. By harnessing engineered dissipations in all-to-all coupled qubits, we demonstrate rapid purification from initially mixed states. Counterintuitively, more mixed initial states reach purified steady states faster, revealing an anomalous relaxation that accelerates both state purification and entanglement generation. We show that the onset of this speedup is governed by the degeneracy of collective subradiant modes, whose dissipative gaps set the purification timescale. Our results provide a platform for exploring anomalous nonequilibrium dynamics in non-Hermitian quantum systems and offer new opportunities for efficient preparation of resource states for quantum information processing.

As a remark, we distinguish several related concepts discussed in this work. Conditional state selection defines the post-selected no-jump dynamics; dissipative filtering selects the asymptotically surviving eigenmode through decay-rate hierarchies; purification reduces linear entropy; coherence generation builds off-diagonal density-matrix elements; and the informational Mpemba effect arises from distinct transient relaxation pathways of different initial states. While some of these concepts are complementary, our informational ME-assisted purification protocol is fundamentally distinct from eigenmode filtering.

Looking forward, more sophisticated relaxation control based on state projection and spectral stretching \cite{Beato2026} may further broaden the scope of facilitated state purification. More intriguingly, the role of engineered long-range interactions \cite{Joshi2024, Chatterjee2024_2} in open or disordered quantum system deserves exploration from the perspective of ME and informational ME \cite{Hsieh2025}. Beyond passive approaches to understanding anomalous relaxation, active Floquet engineering \cite{Yu2026_skin}, projection-based strategy \cite{Gal2020}, or temporary rest channel \cite{Bao2025} may open new avenues for precise control of dynamical behavior in whole or partial quantum systems. Such control could help reveal the underlying mechanisms of subsystem thermalization and offer new insights into related monitored quantum systems \cite{Ashida2020}. 

{\it Acknowledgments}--We acknowledge support from the National Science and Technology Council (NSTC), Taiwan, under the Grants No. 112-2112-M001-079-MY3, No. 115-2112-M-001-035-MY3 and No. 115-2119-M-001-009, and from Academia Sinica under Grant AS-CDA-113-M04. We are also grateful for support from TG 1.2 of NCTS at NTU and helpful discussions with C.-Y. Hsieh. J.-S.Y. acknowledges support from the National Science and Technology Council (NSTC), Taiwan, under Grant No. NSTC 113-2112-M-003-015 - and No. NSTC 114-2112-M-003-005 -, from ``Higher Education Sprout Project`` of National Taiwan Normal University and the Ministry of Education (MOE), Taiwan, and from TG 3.2 of NCTS at NTU.

{\it Data Availability}--The data are not publicly available. The data are available from the authors upon reasonable request. 


\begin{widetext}
 \section{End Matter}
\end{widetext}
 
In this section, we investigate whether the informational Mpemba effect persists under the Lindblad dynamics, which incorporates quantum jumps in addition to the non-Hermitian evolution. These jumps arise from spontaneous emission from the excited state and introduce an additional dissipative channel to the non-Hermitian dynamics considered in the main text. We first investigate how these quantum jumps affect the informational Mpemba effect and compare the trajectory-averaged dynamics with the corresponding master-equation results. We then analyze representative fast and slow trajectories, trajectory-to-trajectory fluctuations, and the distribution of purification times conditioned on monitored quantum-jump events.

\begin{figure*}[!t] 
\centering
\includegraphics[width=\textwidth]{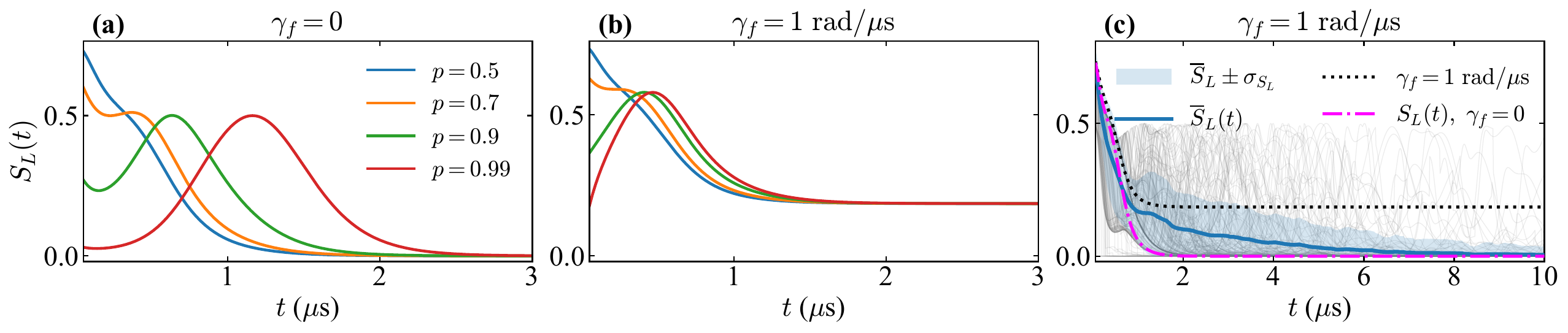}
\includegraphics[width=\textwidth]{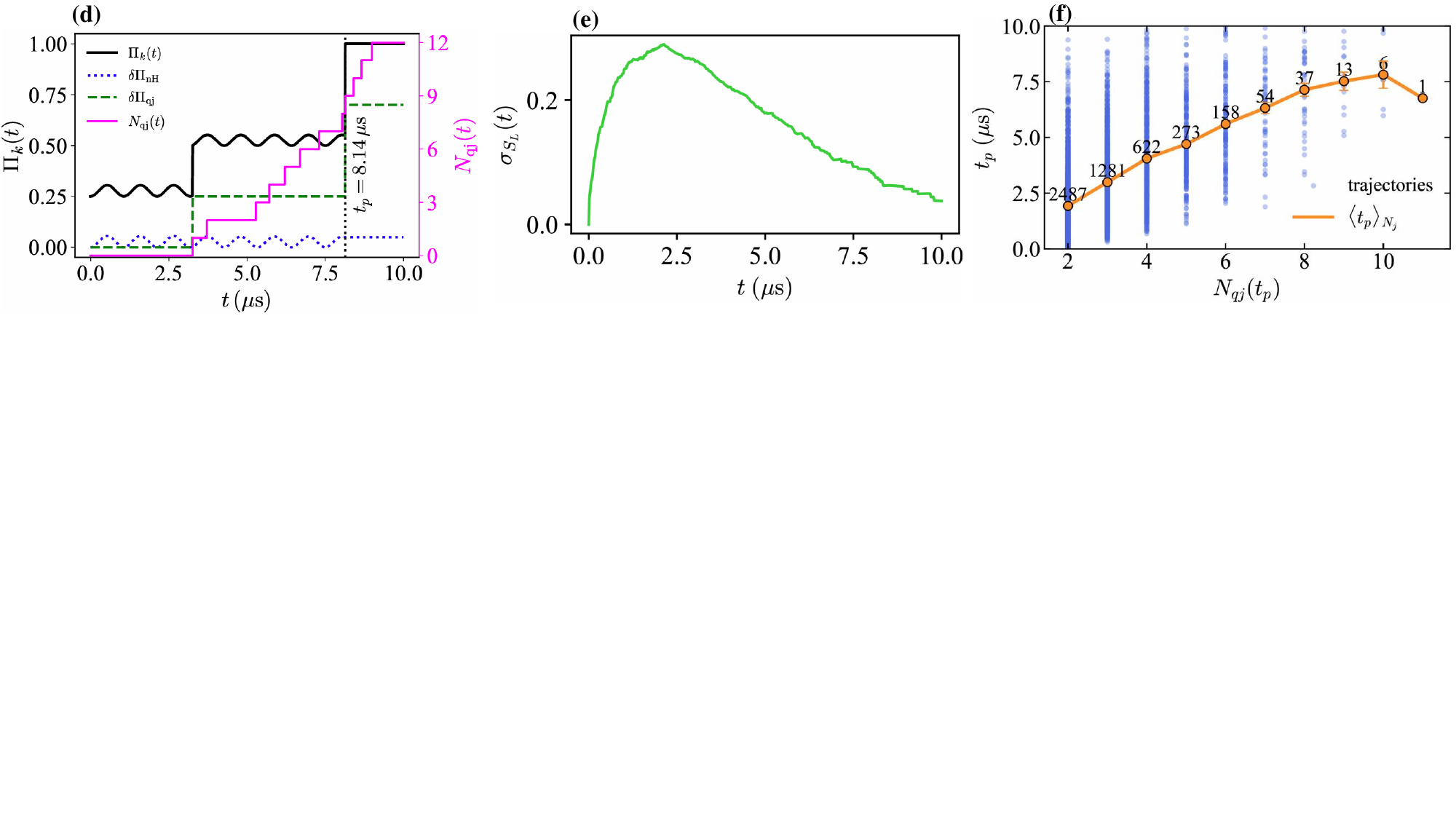} \vspace{-6.75cm}
\caption{Effect of quantum jumps on purification dynamics. Linear entropy under (a) the non-Hermitian dynamics with $\gamma_f=0$ and (b) the master equation with $\gamma_f=1~{\rm rad}/\mu{\rm s}$. (c) Comparison of the master-equation result (black dotted), non-Hermitian dynamics (magenta dash-dotted), and the trajectory average over monitored quantum trajectories (light blue). Thin black curves show 200 individual trajectories, and the gray shaded region around $\overline{S_L}(t)$ indicates the trajectory-to-trajectory fluctuations. (d) Purity of a single quantum trajectory, $\Pi_k(t)$ (black), and its decomposition into the non-Hermitian contribution, $\delta\Pi_{\rm nH}$ (blue dotted), and the quantum-jump contribution, $\delta\Pi_{\rm qj}$ (green dashed). The magenta curve shows the quantum-jump count, $N_{\rm qj}(t)$, along the trajectory. (e) Trajectory-to-trajectory fluctuations quantified by the standard deviation of the linear entropy. (f) First-passage purification times of 5000 quantum trajectories conditioned on the number of quantum jumps, $N_{\rm qj}(t_p)$, accumulated before reaching the threshold purity $\Pi_{\rm th}=0.98$. Blue points denote individual trajectories, while the orange markers show the average $\langle t_p\rangle_{N_{\rm qj}}$ with standard-error bars; the numbers above the markers indicate the number of trajectories in each group. Throughout, we use $\Delta=J=0$, $\eta=1$, and $\gamma_e=6~{\rm rad}/\mu{\rm s}$ in (a)-(c) and $\gamma_e=0$ in (d)-(f).}
\label{qj1} 
\end{figure*}

As discussed in the main text, we restrict our analysis to the effective $\{e,f\}$ manifold, where the local and collective dissipative processes from $|f\rangle$ to $|e\rangle$ modify the effective non-Hermitian dynamics and simultaneously induce quantum jumps. In this End matter, we focus on the local dissipative channel, for which the intrinsic decay associated with each local jump operator $\hat{L}^f_{j}$ is described by the diagonal decay matrix $\Gamma_{jk}^{f}=\gamma_f\delta_{jk}$. In Figs.~\ref{qj1}(a) and (b), we compare the purification dynamics with and without quantum jumps in the same parameter regime considered in the main text, namely $\gamma_e\gg\gamma_f$, with $\gamma_f=1~{\rm rad}/\mu{\rm s}$. We find that states with lower initial purity continue to approach the high-purity steady state faster than initially purer states, demonstrating that the informational Mpemba effect persists in the presence of quantum jumps. Moreover, quantum jumps further accelerate relaxation toward the steady state, with larger jump rates $\gamma_f$ reducing the relaxation time while slightly lowering the steady-state purity, whereas smaller jump rates preserve a higher steady-state purity. Our results therefore demonstrate that the proposed protocol achieves rapid state purification and high steady-state purity under realistic experimental conditions. 

\subsubsection{Trajectory-Level Analysis}
Trajectory-level analysis of continuously monitored open quantum systems provides insight into the role of measurement records in quantum-state purification~\cite{Gopalakrishnan2021}. We therefore employ the quantum-jump method to unravel the Lindblad master equation into individual quantum trajectories~\cite{Wiseman2010,Daley2014}. We initialize each trajectory in $\rho_k(0)=\rho(0)$, which then evolves either under the non-jump non-Hermitian dynamics, $\rho_k(t+dt)=\hat U_{\rm eff}\rho_k(t)\hat U_{\rm eff}^{\dagger}/{\rm Tr}\!\left[\hat U_{\rm eff}\rho_k(t)\hat U_{\rm eff}^{\dagger}\right]$, with $\hat U_{\rm eff}=e^{-i\hat H_{\rm eff}dt}$ and time step $dt$, or undergoes a quantum jump in channel $j$ with probability $dp_j=dt\,{\rm Tr}\!\left[\gamma_f\hat L^{f,\dagger}_{j}\hat L^f_{j}\rho_k(t)\right]$. In the latter case, the state is updated according to $\rho_k(t+dt)=\hat L^f_{j}\rho_k(t)\hat L^{f,\dagger}_{j}/{\rm Tr}\!\left[\hat L^f_{j}\rho_k(t)\hat L^{f,\dagger}_{j}\right]$. Repeating this procedure generates an ensemble of quantum trajectories $\{\rho_k(t)\}$, where $k=1,\ldots,N_{\rm traj}$ labels independent realizations. In the limit $N_{\rm traj}\rightarrow\infty$, averaging over the trajectory ensemble recovers the state obtained from the Lindblad master equation, Eq.~\eqref{rho}. 

We analyze the purification dynamics through the trajectory-resolved linear entropy, $S_{L,k}(t)=1-\Pi_k(t)$ with $\Pi_k(t)={\rm Tr}[\rho_k^2(t)]$, and its trajectory average, $\overline{S_L}(t)=N_{\rm traj}^{-1}\sum_{k=1}^{N_{\rm traj}}S_{L,k}(t)$. Figure~\ref{qj1}(c) compares these results with those obtained from the Lindblad master equation and the non-Hermitian dynamics. Since purity is a nonlinear function of the density matrix, the trajectory-averaged linear entropy differs from that of the Lindblad master equation, i.e., $\overline{S_L}(t)\neq S_L(t)$. This discrepancy arises because the master equation describes the unconditional density matrix obtained by averaging over all possible measurement outcomes, thereby discarding the information contained in individual measurement records and leaving the system in a mixed steady state. By contrast, each quantum trajectory is conditioned on a specific measurement record and therefore purifies, even though the ensemble-averaged density matrix remains mixed. These results demonstrate that quantum-state purification in monitored dynamics of quantum systems is driven by the continuous acquisition of information from the measurement record.

The purification under monitored dynamics is nevertheless slower than the non-Hermitian evolution, in which each quantum trajectory evolves continuously toward the long-lived pure state without quantum jumps. This reduced purification rate arises because trajectories are interrupted by stochastic quantum jumps with different numbers and waiting times, resulting in trajectory-dependent purification rates. These trajectory-to-trajectory fluctuations are evident in Fig.~\ref{qj1}(e) and are quantified by the standard deviation, $\sigma_{S_L}(t)=\sqrt{N_{\rm traj}^{-1}\sum_{k=1}^{N_{\rm traj}}\left[S_{L,k}(t)-\overline{S_L}(t)\right]^2},$ shown as the gray shaded region around $\overline{S_L}(t)$. The fluctuations peak at intermediate times and vanish as all trajectories converge to the same pure steady state. The slow decay of these fluctuations originates from rare quantum trajectories that remain mixed for longer because of the long waiting times between the quantum jumps required for purification. Below, we analyze the role of these rare trajectories in detail.

We next address how purification emerges in the monitored dynamics of the qubits. To this end, we decompose the purity change along each individual quantum trajectory as,
\begin{equation}
\Pi_k(t)=\Pi(0)+\delta\Pi_{{\rm nH},k}(t)+\delta\Pi_{{\rm qj},k}(t),
\end{equation}
where $\Pi(0)=0.25$ is the purity of the maximally mixed initial state, $\delta\Pi_{{\rm nH},k}(t)$ denotes the purity change originating from the non-Hermitian evolution between successive jumps, and $\delta\Pi_{{\rm qj},k}(t)$ denotes the purity change induced by the quantum jumps. To clarify the role of the quantum jumps, we set $\gamma_e=0$, which places the system in the $\mathcal{PT}$-unbroken regime, where the non-Hermitian dynamics does not significantly change the purity apart from weak oscillations around its initial value. The observed purification of the trajectories therefore originates from the quantum jumps, as illustrated in Fig.~\ref{qj1}(d). The trajectory contains 12 quantum jumps during the $10~\mu{\rm s}$ simulation. Despite this large number of jumps, only two contribute to the purification. The first quantum jump increases the purity by $0.25$, after which the non-Hermitian evolution changes the purity only slightly. The next six quantum jumps leave the purity unchanged, and the seventh quantum jump provides the remaining purity increase,
\begin{equation}
\delta\Pi_{{\rm qj}}^{(2)}=1-\Pi(0)-\delta\Pi_{{\rm qj}}^{(1)}-\delta\Pi_{{\rm nH}}(t_2^-),
\end{equation}
required to reach the pure state, where $t_2^-$ denotes the instant immediately before the second quantum jump. Once the trajectory becomes pure, neither subsequent quantum jumps nor the non-Hermitian evolution alter its purity. 

To gain further insight into the role of quantum jumps in the purification process, we examine the dependence of the first-passage purification time on the number of quantum jumps~\cite{Menczel2026}. For each trajectory $k$, the first-passage purification time is defined as the first instant at which the trajectory purity exceeds the threshold $\Pi_{\rm th}=0.98$, namely $t_p^{(k)}=\inf{t>0:\Pi_k(t)\ge\Pi_{\rm th}}$. The threshold is chosen because the effective quantum jumps always increase the purity, ensuring that every trajectory reaching $\Pi_{\rm th}$ subsequently evolves to the pure steady state. We then group trajectories according to the number of quantum jumps $N_{\rm qj}(t_p)$ recorded before reaching the threshold purity $\Pi_{\rm th}$. Figure~\ref{qj1}(f) shows the resulting dependence of $t_p^{(k)}$ on $N_{\rm qj}(t_p)$. Each blue point represents a single trajectory, while the orange markers denote the average, $\langle t_p\rangle_{N_{\rm qj}}=N_{N_{\rm qj}}^{-1}\sum_{k\in N_{\rm qj}}t_p^{(k)}$, where $N_{N_{\rm qj}}$ is the number of trajectories with the same jump count.

Figure~\ref{qj1}(f) reveals nearly half of the 5000 trajectories reach the purification threshold after two quantum jumps, whereas trajectories with larger jump counts require longer times to purify and become rare. These rare trajectories sustain the late-time trajectory-to-trajectory fluctuations, thereby delaying the trajectory-averaged purification relative to the no-jump non-Hermitian dynamics. The finite vertical spread at each value of $N_{\rm qj}$ further shows that trajectories with the same jump count can purify at different times because of the stochastic waiting times between quantum jumps. Since two effective quantum jumps are required for purification, trajectories with $N_{\rm qj}>2$ necessarily contain ineffective quantum jumps that occur before the second effective jump without changing the purity. Thus, quantum jumps either increase the purity or leave it unchanged, and the stochastic timing of those that increase the purity determines the purification dynamics of the monitored quantum system. 

\end{document}